\documentclass[10pt,aps,pra,showpacs,twocolumn]{revtex4}
\usepackage{amsmath}
\usepackage{amssymb}

\usepackage{amsmath,amssymb}
\usepackage[usenames]{color}
\usepackage{mathbbol}
\usepackage{amssymb}
\usepackage{grffile}
\usepackage[pdftex]{graphicx}
\usepackage{amsmath, amstext, amssymb, amsfonts, amsxtra}
\usepackage{textcomp}
\usepackage{xspace}

\DeclareSymbolFontAlphabet{\amsmathbb}{AMSb}

\newcommand{\aaop}{\hat{\alpha}}
\newcommand{\aadop}{\hat{\alpha}^{\dagger}}
\newcommand{\aop}{\hat{a}}
\newcommand{\adop}{\hat{a}^{\dagger}}

\newcommand{\bop}{\hat{b}}
\newcommand{\bdop}{\hat{b}^{\prime}}

\newcommand{\Hop}{\hat{H}}
\newcommand{\rhoop}{\hat{\rho}}
\newcommand{\hm}{\bold{h}}
\newcommand{\Mm}{\bold{M}}
\newcommand{\Pm}{\bold{P}}
\newcommand{\Am}{\bold{A}}
\newcommand{\Om}{\bold{O}}
\newcommand{\IM}{\bold{I}}
\newcommand{\Jm}{\bold{J}}
\newcommand{\Cm}{\bold{C}}
\newcommand{\Dm}{\bold{D}}
\newcommand{\Um}{\bold{U}}
\newcommand{\Vm}{\bold{V}}
\newcommand{\Ym}{\bold{Y}}
\newcommand{\Zm}{\bold{Z}}
\newcommand{\Wm}{\bold{W}}
\newcommand{\Qm}{\bold{Q}}
\newcommand{\Km}{\bold{K}}
\newcommand{\Xm}{\bold{X}}
\newcommand{\Op}{\hat{O}}
\newcommand{\Sop}{\hat{S}}
\newcommand{\Top}{\hat{T}}

\newcommand{\Dop}{\mathcal{D}}
\newcommand{\Jp}{J^{\perp}}
\newcommand{\Jpa}{J^{\parallel}}

\newcommand{\Lmp}{\bold{\Lambda}^{+}}
\newcommand{\Lmm}{\bold{\Lambda}^{-}}

\newcommand{\Omegam}{\bold{\Omega}}
\newcommand{\deltam}{\bold{\delta}}
\newcommand{\id}{\bold{1}} 
\newcommand{\zero}{\bold{0}}
\newcommand{\im}{{\rm i}}   
\newcommand{\real}{\mathcal{R}}
\newcommand{\imag}{\mathcal{I}}
\newcommand{\hc}{{\rm H.c.}} 
\newcommand{\diag}{{\rm diag}}
\newcommand{\tr}{{\rm tr}}

\newcommand{\stackvert}[2]{\genfrac{(}{)}{0pt}{}{#1}{#2}}
\newcommand{\lmp}{{\pmb{\lambda}_P}} 
\newcommand{\lmpc}{{\pmb{\lambda}_P^{\ast}}} 
\newcommand{\pare}[1]{\left(#1 \right)}
\newcommand{\spare}[1]{\left[#1 \right]}

\newcommand{\sutd}{Singapore University of Technology and Design, 8 Somapah Road, 487372 Singapore}

\begin{document}

\title{Solutions for dissipative quadratic open systems: part I - bosons}  
\author{Chu Guo}
\affiliation{\sutd} 
\author{Dario Poletti}
\affiliation{\sutd} 

\begin{abstract} 
This is a work in two parts in which we show how to solve a large class of Lindblad master equations for non-interacting particles on $L$ sites. In part I we concentrate on bosonic particles. We show how to reduce the problem to diagonalizing an $L \times L$ non-Hermitian matrix. In particular, for boundary dissipative driving of a uniform chain, the matrix is a tridiagonal bordered Toeplitz matrix which can be solved analytically for the normal master modes and their relaxation rates (rapidities). In the regimes in which an analytical solution cannot be found, our approach can still provide a speed-up in the numerical evaluation. We use this numerical method to study the relaxation gap at non-equilibrium phase transitions in a boundary driven bosonic ladder with synthetic gauge fields. We conclude by showing how to construct the non-equilibrium steady state. 
\end{abstract}

\date{\today}
\pacs{03.65.Yz, 05.60.Gg, 05.30.Jp}
\maketitle


\section{Introduction} 

Quantum systems in contact with an environment display a very rich physics including emergence of non-equilibrium phase transitions \cite{DiehlZoller2008, DiehlZoller2010, DallaTorreAltman2010} which can be used to engineer particularly interesting non-equilibrium steady states (NESS). 
Moreover, also the relaxation towards an asymptotic or steady state can, for example, manifest a non-trivial dynamics, from power-law to stretched exponentials and even aging \cite{PolettiKollath2012, CaiBarthel2012, PolettiKollath2013, SciollaKollath2015}.  
A particularly important class of open quantum system is that of boundary driven systems, in which a current may be induced by the coupling, only at the extremities, of the system to the environment. This class of systems is of particular relevance in the study of quantum transport.     

The knowledge of analytical solutions for open quantum systems would allow to build a better intuition of the physics of these systems and to test numerical methods. 
The NESS of a boundary driven Heisenberg model can, in many regimes, be computed analytically using a Matrix Product Ansatz as shown in \cite{Prosen2011a, Prosen2011b, Prosen2014, Popkov2013b, Popkov2014, MedvedyevaProsen2016} (for a review see \cite{Prosen2015}).     
In \cite{Znidaric2010} an exact solution for a diffusive XX chain is presented using a cleverly designed ansatz. For a boundary driven bosonic non-interacting system, in \cite{AsadianBriegel2012} the authors showed how to analytically compute the local densities and the current.   
In the seminal article \cite{Prosen2008}, which is particularly relevant to our work, Prosen showed that for a quadratic open fermionic model with $L$ sites, solving for the relaxation rates of the quantum Lindblad equation, can be reduced to the diagonalization of a $4L \times 4L$ anti-symmetric matrix, which can be further reduced to the diagonalization of a $2L \times 2L$ general matrix \cite{Prosen2010}. A similar method was also applied to quadratic open bosonic model \cite{Prosen2010b}. 

Here we build on this strategy while focusing on a boundary dissipative driven quadratic bosonic system, whose Hamiltonian conserves the total particle number. In the second part of this work we will consider the case of fermions. We are able to reduce the problem to the diagonalization of a $L \times L$ matrix which can thus be studied numerically more effectively. Moreover we show that in many physically relevant cases the matrix to be diagonalized is a tridiagonal bordered Toeplitz matrix for which analytical expressions for the eigenvalues and eigenvectors are known. We also show an example for which we can explicitly write the relaxation rates (rapidities) of all the normal master modes of the Lindblad master equation. This can be used, for example, to compute the relaxation gap, that is the rapidity of the slowest decaying normal master mode. Then we use this method to study the scaling of the relaxation gap in a system with two phase transitions of different nature. We show that the scaling of the relaxation gap is different in the two cases. Last we use the framework developed to give an expression for the steady state.         

This paper is organized as follows: In section \ref{sec:model} we introduce the quadratic bosonic model we study. In section \ref{sec:diagonalization}, we show how to diagonalize the Lindblad master equation and obtain the normal master modes. In section \ref{sec:exact} we show how to solve analytically the boundary driven bosonic chain and study the scaling of the relaxation gap. In section \ref{sec:algorithm}, we introduce an efficient numerical algorithm to compute the quadratic observables such as currents and densities. Then we use it to study the relaxation gap of a system which presents non-equilibrium phase transitions. In section \ref{sec:steady_state} we analytically construct a similarity transformation which maps the vacuum state to the steady state of the system and in section \ref{sec:conclusion}, we draw our conclusions.

\section{model} \label{sec:model}
We consider an open quantum systems of $L$ sites with bosonic particles. Its dynamics is described by the quantum Lindblad master equation \cite{GoriniSudarshan1976, Lindblad1976} with Lindbladian $\mathcal{L}$
\begin{align}
\frac{d}{dt} \rhoop = \mathcal{L}(\rhoop) = -\frac{\im}{\hbar} [ \Hop, \rhoop ] + \mathcal{D}(\rhoop). \label{eq:first}
\end{align}
Here $\rhoop$ is the density operator of the system, $\Hop$ is the Hamiltonian, and $\mathcal{D}$, the dissipator, describes the dissipative part of the evolution. The Hamiltonian $\hat{H}$ is given by 
\begin{align}
\Hop =\sum_{m,n=1}^{L}\hm_{m,n}\aadop_m \aaop_n,
\end{align}
where $\hm$ is an $L\times L$ Hermitian matrix, and $\aadop_j$($\aaop_j$) creates(annihilates) one boson on site $j$. The dissipative part is given by
\begin{align}
\mathcal{D}(\rhoop) = \sum_{i,j=1}^{L}&\left[\Lmp_{i,j}(\aadop_{i}\rhoop \aaop_{j}-\aaop_{j}\aadop_{i}\rhoop) \right. \\ &+ \left.  \Lmm_{i,j}(\aaop_{i}\rhoop \aadop_{j}-\aadop_{j}\aaop_{i}\rhoop)+ \hc \right],
\end{align}
where $\Lmp$ and $\Lmm$ are $L \times L$ Hermitian and non-negative matrices. In the trivial case of only one site, the dissipator has the familiar form $\mathcal{D}(\rhoop) = \Lambda^+(\aadop\rhoop \aaop-\aaop\aadop\rhoop) + \Lambda^-(\aaop\rhoop \aadop-\aadop\aaop\rhoop) + \hc$, where $\Lambda^{\pm}$ are the heating/cooling rates. In general, one can make a unitary transformation to the creation and annihilation operators so that $\Lmp$ and $\Lmm$ become diagonal. We note that while the Hamiltonian that we consider conserves the total quantum number, the dissipative part in general does not.

\section{solving the master equation} \label{sec:diagonalization}

\subsection{Reshaping the density operator in a new representation}

We perform a one-to-one mapping from the density operator basis elements $\vert n_1, n_2, \dots n_L \rangle \langle  n'_1, n'_2, \dots n'_L \vert $ to a state vector basis (with $2L$ sites) $\vert n_1, \dots n_L, n'_1, \dots n'_{L} \rangle$ (see for example \cite{ProsenPizorn2008, ProsenZunkovic2010, Prosen2010, PizornTroyer2013}). From this, the operator $\aaop_i$ acting on site $i$ on the left of the density matrix is mapped to $\aop_i$ acting on the state vector on the $i$-th site, while the operator $\aaop_i$ acting on the right of the density matrix is mapped to $\adop_{L+i}$ acting on the state vector. We denote the density operator to be $\vert \rho \rangle$ in the new representation.

\subsection{The master equation in the new representation}

$\mathcal{L}$ in Eq.(\ref{eq:first}) can thus be written as     
 \begin{eqnarray}
\mathcal{L} &&= \left(
                                                         \begin{array}{cc}
                                                          \textbf{a}_{1\rightarrow L}^{\dagger} \\
                                                          \textbf{a}_{L+1\rightarrow 2L} \\
                                                         \end{array}
                                                       \right)^t \Mm
             \left(
                                                         \begin{array}{cc}
                                                          \textbf{a}_{1\rightarrow L} \\
                                                          \textbf{a}_{L+1\rightarrow 2L}^{\dagger} \\
                                                         \end{array}
                                                       \right) \nonumber \\
&&+ \left(
                                                         \begin{array}{cc}
                                                          \textbf{a}_{1\rightarrow L} \\
                                                          \textbf{a}_{L+1\rightarrow 2L}^{\dagger} \\
                                                         \end{array}
                                                       \right)^t \Mm^t
             \left(
                                                         \begin{array}{cc}
                                                          \textbf{a}_{1\rightarrow L}^{\dagger} \\
                                                          \textbf{a}_{L+1\rightarrow 2L} \\
                                                         \end{array}
                                                       \right) \nonumber \\
                                             &&+ \tr({\Lmm}^t-\Lmp)    \label{eq:master}
\end{eqnarray}
where $\Mm$ is a $2L \times 2L$ matrix,
\begin{eqnarray}
\Mm = \left(
                                                         \begin{array}{cc}
                                                          \Km & \Lmp  \\
                                                          {\Lmm}^t & \Km^{\dagger} \\
                                                         \end{array}
                                                       \right).  \label{eq:Mm}
\end{eqnarray}
Here $\Km = (-\im \hm/\hbar -\Lmp - {\Lmm}^t)/2$, where with $\Am^t$ we indicate the transpose of the matrix $\Am$. We have also used the notation $\textbf{a}_{1\rightarrow L} $ to denote the column vector made of operators $\aop_1, \aop_2 ,\dots,\aop_L$  and $\textbf{a}^{\dagger}_{1\rightarrow L} $ for the column vector made of $\adop_{1}, \adop_{2} ,\dots,\adop_{L}$, the same applies for $\textbf{a}_{L+1\rightarrow 2L} $ and $\textbf{a}^{\dagger}_{L+1\rightarrow 2L} $. Here we stress that the Liouvillian $\mathcal{L}$ can be written in the simple form of Eq.(\ref{eq:master}) because we study a number conserving Hamiltonian. If the Hamiltonian is not number conserving, the coefficient matrix will be a $4L \times 4L$ matrix which can not be written in block diagonal form. 

\subsection{Normal master modes of the master equation}

In general $\Mm$ is not Hermitian and it cannot always be diagonalized, however in the following we start from the assumption that we know a transformation which can diagonalize $\Mm$ and preserves bosonic commutation relations. This assumption is a posteriori verified in all the cases we considered. This transformation is given by the matrices $\Wm_1$ and $\Wm_2$ as follows  
\begin{eqnarray}
 &&\left(
                                  \begin{array}{cc}
                                  \textbf{a}_{1\rightarrow L} \\
                                 \textbf{a}_{L+1\rightarrow 2L}^{\dagger} \\
                                  \end{array}
                                  \right) = \Wm_1\left(
                                  \begin{array}{cc}
                                  \textbf{b}_{1\rightarrow L} \\
                                  \textbf{b}_{L+1\rightarrow 2L}^{\prime} \\
                                   \end{array} \right) \\
                                  &&\left(
                                  \begin{array}{cc}
                                 \textbf{a}_{1\rightarrow L}^{\dagger} \\
                                 \textbf{a}_{L+1\rightarrow 2L} \\
                                  \end{array}
                                  \right) = \Wm_2\left(
                                  \begin{array}{cc}
                                  \textbf{b}_{1\rightarrow L}^{\prime}  \\
                                 \textbf{b}_{L+1\rightarrow 2L}\\
                                 \end{array} \right)                                                        
\end{eqnarray} 
where as for $\textbf{a}_{1\rightarrow L}$ and $\textbf{a}^{\dagger}_{1\rightarrow L}$, $\textbf{b}_{1\rightarrow L}$ means the column vector made of operators $\bop_1, \bop_2 ,\dots,\bop_L$ and $\textbf{b}^{\prime}_{1\rightarrow L}$ means the column vector made of $\bdop_{1}, \bdop_{2} ,\dots,\bdop_{L}$. Again the same notation applies for $\textbf{b}_{L+1\rightarrow 2L}$ and $\textbf{b}^{\prime}_{L+1\rightarrow 2L}$.

Using this transformation we get   
 \begin{eqnarray}
\mathcal{L} &&= \left(
             \begin{array}{cc}
              \textbf{b}_{1\rightarrow L}^{\prime} \\
              \textbf{b}_{L+1\rightarrow 2L} \\
             \end{array}
             \right)^t \Wm_2^{t} \Mm \Wm_1
             \left(
              \begin{array}{cc}
              \textbf{b}_{1\rightarrow L} \\
           \textbf{b}_{L+1\rightarrow 2L}^{\prime} \\
             \end{array}
              \right) \nonumber \\
&& + \left(
              \begin{array}{cc}
              \textbf{b}_{1\rightarrow L} \\
           \textbf{b}_{L+1\rightarrow 2L}^{\prime} \\
             \end{array}
              \right)^t \Wm_1^{t} \Mm^t \Wm_2
             \left(
             \begin{array}{cc}
              \textbf{b}_{1\rightarrow L}^{\prime} \\
              \textbf{b}_{L+1\rightarrow 2L} \\
             \end{array}
             \right) \nonumber \\
             &&+ \tr({\Lmm}^t-\Lmp)
\end{eqnarray}
The bosonic commutation relation can be written as $$\left[\left(
               \begin{array}{cc}
              \textbf{a}_{1\rightarrow L} \\
              \textbf{a}_{L+1\rightarrow 2L}^{\dagger} \\
              \end{array}
              \right), \left(
               \begin{array}{cc}
              \textbf{a}_{1\rightarrow L}^{\dagger} \\
             \textbf{a}_{L+1\rightarrow 2L} \\
             \end{array}
            \right)^{t}\right] = \Zm_L,$$ and requiring for the bosonic commutation relation to apply also to the $\bop$ we get  
            $$\left[\left(
             \begin{array}{cc}
             \textbf{b}_{1\rightarrow L} \\
            \textbf{b}_{L+1\rightarrow 2L}^{\prime} \\
             \end{array}
             \right), \left(
            \begin{array}{cc}
           \textbf{b}_{1\rightarrow L}^{\prime} \\
             \textbf{b}_{L+1\rightarrow 2L} \\
             \end{array}
             \right)^{t}\right] = \Zm_L$$ and hence
\begin{eqnarray}
\Zm_{L} &=& \Wm_1\Zm_{L}\Wm_2^t\;
\Longleftrightarrow \; \Wm_2 = \Zm_{L} {\Wm_1^{t}}^{-1} \Zm_{L}
\end{eqnarray}
Here we have used 
\begin{align} \label{eq:eqZ}
\Zm_{L}=\left(
                                                         \begin{array}{cc}
                                                         \id_{L}  & 0 \\
                                                          0 & -\id_{L} \\
                                                         \end{array}
                                                       \right)
\end{align}
where we have denoted $\id_{l}$ for an identity matrix of size $l$. In the following we will also use the matrices 
\begin{align}\label{eq:eqX}
\Xm_{L}=\left(
                                                         \begin{array}{cc}
                                                          0 & \id_{L} \\
                                                          \id_{L} & 0 \\
                                                         \end{array}
                                                       \right)
\end{align}
and 
\begin{align}\label{eq:eqY}
\Ym_{L}= -\im \left(
                                                         \begin{array}{cc}
                                                          0 & \id_{L} \\
                                                          -\id_{L} & 0 \\
                                                         \end{array}
                                                       \right).
\end{align}
Matrices in Eqs.(\ref{eq:eqZ}, \ref{eq:eqX}, \ref{eq:eqY}), being given by a tensor product between Pauli matrices and identity, satisfy the relations
\begin{align} 
&\Zm_{L}^2=\id_{2L}, \;\;\Xm_{L}^2=\id_{2L}, \label{eq:xandz}\\ 
&\Ym_{L}^2 = \id_{2L}, \;\;\Zm_{L}\Xm_{L}=-\Xm_{L}\Zm_{L}= \im\Ym_{L}.  \label{eq:y}
\end{align}

It follows that 
  \begin{align}
\mathcal{L} &= \stackvert{\textbf{b}_{1\rightarrow L}^{\prime}}{\textbf{b}_{L+1\rightarrow 2L} }^t \Zm_{L} \Wm_1^{-1}\Zm_{L} \Mm \Wm_1   \stackvert{\textbf{b}_{1\rightarrow L} } {\textbf{b}_{L+1\rightarrow 2L}^{\prime} } \nonumber \\ 
&+ \stackvert{\textbf{b}_{1\rightarrow L} } {\textbf{b}_{L+1\rightarrow 2L}^{\prime} }^t \Wm_1^{t}\Mm^t \Zm_{L} {\Wm_1^{t}}^{-1} \Zm_L \stackvert{\textbf{b}_{1\rightarrow L}^{\prime}}{\textbf{b}_{L+1\rightarrow 2L}}\nonumber \\
&+\tr({\Lmm}^t-\Lmp) \nonumber \\
&= \stackvert{\textbf{b}_{1\rightarrow L}^{\prime}}{\textbf{b}_{L+1\rightarrow 2L} }^t \Zm_{L} (\Wm_1^{-1}\Zm_{L} \Mm \Wm_1) \stackvert{\textbf{b}_{1\rightarrow L} }{ \textbf{b}_{L+1\rightarrow 2L}^{\prime} } \nonumber \\
&+  \stackvert{\textbf{b}_{1\rightarrow L} } {\textbf{b}_{L+1\rightarrow 2L}^{\prime} }^t (\Wm_1^{-1}\Zm_L \Mm \Wm_1)^t \Zm_L \stackvert{\textbf{b}_{1\rightarrow L}^{\prime}}{\textbf{b}_{L+1\rightarrow 2L} } \nonumber \\
&+ \tr({\Lmm}^t-\Lmp). 
\end{align}
This implies that the problem of finding the normal modes of the system reduces to finding a $\Wm_1$ such that $\Zm_L \Mm$ can be diagonalized, that is, 
\begin{eqnarray}
\Wm_1^{-1}(\Zm_L \Mm) \Wm_1 = \diag(\beta_1, \beta_2, \dots, \beta_{2L}), \label{eq:eigeneq}
\end{eqnarray}
where $\diag(\vec{v})$ is a diagonal matrix with the elements of the vector $\vec{v}$ on its diagonal. It would then be possible to write the following compact form for $\mathcal{L}$:    
 \begin{align}
\mathcal{L} &= 2\sum_{i=1}^{L}(\beta_{i}\bdop_{i}\bop_{i} - \beta_{L+i}\bdop_{L+i}\bop_{L+i}) \nonumber \\
 &+\sum_{i=1}^{L}(\beta_{i}-\beta_{L+i})+\tr({\Lmm}^t-\Lmp) \label{eq:me_beta}
\end{align}

\subsection{Diagonalizing $\Zm_L \Mm $} \label{sec:diagonalizationZM}

Here we will explicitly construct the eigenvalues and eigenvectors of the matrix $\Zm_L \Mm$. From Eq.(\ref{eq:Mm}) we notice that $\Mm$ satisfies the relation 
\begin{align}
\Xm_L \Mm \Xm_L = \Mm^{\dagger}.
\end{align} 
Therefore, if $x = \left(
                                                         \begin{array}{cc}
                                                          \textbf{u} \\
                                                          \textbf{v} \\
                                                         \end{array}
                                                       \right)$ is a right eigenvector of $\Zm_{L} \Mm$ with eigenvalue $\omega$, then $x^{\dagger}\Ym_L$ is a left eigenvector of $\Zm_{L} \Mm$ with eigenvalue $-\omega^{\ast}$.
In fact, using Eq.(\ref{eq:xandz})   
\begin{align}
&\Zm_L \Mm x = \omega x \rightarrow x^{\dagger}\Mm^{\dagger}\Zm_{L}= \omega^{\ast}x^{\dagger} \nonumber \\ 
&\rightarrow
x^{\dagger} \Xm_{L} \Xm_{L} \Mm^{\dagger} \Xm_{L} = \omega^{\ast}x^{\dagger} \Zm_{L} \Xm_{L} \nonumber \\ 
&\rightarrow  x^{\dagger} \Xm_{L} \Mm  = \omega^{\ast}x^{\dagger} \Zm_{L} \Xm_{L}. \nonumber                
\end{align}
 This implies, using Eq.(\ref{eq:y}), that $x^{\dagger}\Xm_{L}\Zm_{L} \Zm_{L} \Mm = x^{\dagger} \Xm_{L} \Mm = \omega^{\ast}x^{\dagger} \Zm_{L} \Xm_{L}$                                                    , which means $x^{\dagger}\Ym_{L} \Zm_{L} \Mm = -\omega^{\ast}x^{\dagger}\Ym_{L}$. Thus it follows that 
\begin{align}
&x^{\dagger}\Ym_{L} \left(\Zm_{L} \Mm\right)=-\omega^{\ast} x^{\dagger}\Ym_{L} \label{eq:lefteigen}
\end{align} 
i.e. $x^{\dagger}\Ym_{L}$ is a left eigenvector of $\Zm_{L} \Mm$. 
                                                         
Moreover if $x_1$ is a right eigenvector of $\Zm_{L} \Mm$ with eigenvalue $\omega_1$, and $x_2$ is a right eigenvector of $\Zm_{L}\Mm$ with eigenvalue $\omega_2$, then if $\omega_1+\omega_2^{\ast} \neq 0$ it follows that $x_1^{\dagger}\Ym_L x_2 = 0$.
In fact 
\begin{align}
&\Zm_{L} \Mm x_1 = \omega_1 x_1; \nonumber \\
&\Zm_{L} \Mm x_2 = \omega_2 x_2, \nonumber 
\end{align}
then 
\begin{align}
&x_1^{\dagger}\Ym_{L}\Zm_{L} \Mm = -\omega_1^{\ast} x_1^{\dagger}\Ym_{L}; \nonumber \\
&\Zm_{L} \Mm x_2 = \omega_2 x_2; \nonumber \\
&\rightarrow (\omega_1^{\ast}+\omega_2)x_1^{\dagger}\Ym_{L} x_2 = 0 \nonumber
\end{align}

Since the eigenvalues of $\Zm_L \Mm$ always appear in pairs, we could list the eigenvalues and the corresponding eigenvectors of $\Zm_L \Mm$ as $\omega_1, \omega_2, \dots, \omega_L, -\omega_{1}^{\ast},\dots, \omega_L^{\ast}$, with the matrix $\Wm_1$ composed in each column by the right eigenvectors $\Wm_1=(\vec{x}_1, \vec{x}_2,\dots, \vec{x}_{2L})$. Then following Eq.(\ref{eq:lefteigen}) we know that $\vec{x}^{\dagger}_{L+j} \Ym_L$ is the left eigenvector of $\Zm_L \Mm$ correponds to $\omega_j$, and $\vec{x}^{\dagger}_j \Ym_L$ is the left eigenvector corresponds to $-\omega_j^{\ast}$, for $1 \leq j \leq L$. Therefore the left eigenvectors of $\Zm_L \Mm$ constitute the matrix $\Xm_L \Wm_1^{\dagger} \Ym_L$. We can now choose to renormalize the right eigenvectors as 
\begin{eqnarray}
\im\Xm_L \Wm_1^{\dagger} \Ym_L \Wm_1 = \Zm_L \Leftrightarrow  \Ym_L \Wm_1^{\dagger} \Ym_L \Wm_1 = -\id_{2L}
\end{eqnarray}
so that we have
\begin{eqnarray}
\Wm_1^{-1} &=& -\Ym_L \Wm_1^{\dagger} \Ym_L, \\
\Wm_2 &=& -\Xm_L \Wm_1^{\ast} \Xm_L.
\end{eqnarray}

At this point we define a new $L\times L$ matrix $\Pm$, which satisfies
\begin{eqnarray} \label{eq:defineP}
\Pm = \Km + \Lmp = (-\im \hm/\hbar +\Lmp - {\Lmm}^t)/2,
\end{eqnarray}
for which we assume to have the eigendecomposition
\begin{align} \label{eq:eigenP}
\Pm \Wm_P = \Wm_P \lmp,
\end{align}
 where $\Wm_P$ and $\lmp$ are eigenvectors and eigenvalues. Then we find that the $2L\times L$ matrix formed by $\stackvert{\Wm_P}{\Wm_P}$ constitutes $L$ right eigenvectors of $\Zm_L \Mm$, corresponding to $\lmp$, and the $L\times 2L$ matrix $(\Wm_P^{\dagger}\; ,\; -\Wm_P^{\dagger})$ constitutes $L$ left eigenvectors of $\Zm_L \Mm$, corresponding to $-\lmpc$. 
This can be shown from      
\begin{align}
\Zm_L \Mm \stackvert{\Wm_P}{\Wm_P} &= \stackvert{\Pm \Wm_P}{\Pm \Wm_P} = \stackvert{\Wm_P}{\Wm_P}\lmp \nonumber 
\end{align} 
and 
\begin{align}  
( \Wm_P^{\dagger}, -\Wm_P^{\dagger} )\Zm_L \Mm &= (-\Wm_P^{\dagger} \Pm^{\dagger}, \Wm_P^{\dagger} \Pm^{\dagger}) \nonumber \\ 
&= -\lmpc(\Wm_P^{\dagger}, -\Wm_P^{\dagger}) \nonumber
\end{align} 
Hence by denoting the remaining $L$ right eigenvectors of $\Zm_L \Mm$ as $\stackvert{\Cm}{\Dm}$, where $\Cm,$ $\Dm$ are $L\times L$ matrices, we know that they form the right eigenvectors with eigenvalues $-\lmpc$, which are paired with the left eigenvectors $(\Wm_P^{\dagger}\; -\Wm_P^{\dagger})$. Also $(-\Dm^{\dagger}\; \Cm^{\dagger})$ will be the left eigenvectors corresponding the eigenvalues $\lmp$, which are paired with the right eigenvectors $\stackvert{\Wm_P}{\Wm_P}$. 

Therefore $\Wm_1$ and $\Wm_2$ can be written more explicitly as
\begin{align}
\Wm_1 &= \left(
             \begin{array}{cccc}
              \Wm_P & \Cm  \\
              \Wm_P & \Dm \\
              \end{array}
         \right), \;\;          \Wm_1^{-1} = \left(
             \begin{array}{cccc}
              -\Dm^{\dagger} & \Cm^{\dagger}  \\
              \Wm_P^{\dagger} & -\Wm_P^{\dagger} \\
              \end{array}
         \right) ,   \label{eq:W1} \\
\Wm_2 &= -\left(
              \begin{array}{cccc}
              \Dm^{\ast} & \Wm_P^{\ast}  \\
              \Cm^{\ast} & \Wm_P^{\ast} \\
              \end{array}
          \right) ,  \;\; \Wm_2^{-1} = \left(
              \begin{array}{cccc}
              \Wm_P^t & -\Wm_P^t  \\
              -\Cm^t & \Dm^t \\
              \end{array}
          \right) \label{eq:W2}                                                         
\end{align} 
which means
\begin{subequations}
\begin{align}
\textbf{a}_{1 \rightarrow L} &= \Wm_P \textbf{b}_{1 \rightarrow L} + \Cm \textbf{b}_{L+1 \rightarrow 2L}^{\prime}; \\
\textbf{a}_{L+1 \rightarrow 2L}^{\dagger} &= \Wm_P \textbf{b}_{1 \rightarrow L} + \Dm \textbf{b}_{L+1 \rightarrow 2L}^{\prime}; \\
\textbf{a}_{1 \rightarrow L}^{\dagger} &= -\Dm^{\ast} \textbf{b}_{1 \rightarrow L}^{\prime} - \Wm_P^{\ast} \textbf{b}_{L+1 \rightarrow 2L}; \\
\textbf{a}_{L+1 \rightarrow 2L} &= -\Cm^{\ast} \textbf{b}_{1 \rightarrow L}^{\prime} - \Wm_P^{\ast} \textbf{b}_{L+1 \rightarrow 2L}
\end{align} \label{eq:atob}
\end{subequations}
and the inverse equation
\begin{subequations} 
\begin{align}
\textbf{b}_{1 \rightarrow L} &= -\Dm^{\dagger} \textbf{a}_{1 \rightarrow L} + \Cm^{\dagger} \textbf{a}_{L+1 \rightarrow 2L}^{\dagger}; \\
\textbf{b}_{L+1 \rightarrow 2L}^{\prime} &= \Wm_P^{\dagger} \textbf{a}_{1 \rightarrow L} - \Wm_P^{\dagger} \textbf{a}_{L+1 \rightarrow 2L}^{\dagger}; \\
 \textbf{b}_{1 \rightarrow L}^{\prime} &= \Wm_P^{t} \textbf{a}_{1 \rightarrow L}^{\dagger} - \Wm_P^{t} \textbf{a}_{L+1 \rightarrow 2L}; \\
  \textbf{b}_{L+1 \rightarrow 2L} &= -\Cm^{t} \textbf{a}_{1 \rightarrow L}^{\dagger} + \Dm^{t} \textbf{a}_{L+1 \rightarrow 2L}
\end{align} \label{eq:btoa}    
\end{subequations}
Noticing that $\sum \lambda_{P,i} = \tr(\Pm) = \left[-\im\;\tr(\hm/\hbar) - \tr({\Lmm}^t-\Lmp) \right]/2$ and since the $(\lambda_{P,1}, \dots \lambda_{P,L},\; -\lambda_{P,1}^{\ast}, \dots -\lambda_{P,L}^{\ast})$ correspond to $(\beta_{1},\dots \beta_{ 2L})$, the eigenvalues of $\Zm_L \Mm$, we get the following identity           
\begin{align}
\sum_{i=1}^L(\beta_i - \beta_{L+i}) = \sum_{i=1}^L(\lambda_{P,i} + \lambda_{P,i}^{\ast}) = \tr(\Lmp-{\Lmm}^t) \label{eq:summingrule}
\end{align}
which exactly cancels the last term in the expression of $\mathcal{L}$ in Eq.(\ref{eq:me_beta}). 
 
We can then write $\mathcal{L}$ as
\begin{align}
\mathcal{L} = 2\sum_{i=1}^{L}\lambda_{P,i}\bdop_{i}\bop_{i} + 2\sum_{i=1}^L \lambda_{P,i}^{\ast}\bdop_{L+i}\bop_{L+i}. \label{eq:Llambdap}
\end{align}     
The state $|\rho_{ss}\rangle$ which annihilates all the operator $\textbf{b}_{1 \rightarrow 2L}$ is the steady state because $\mathcal{L} |\rho_{ss}\rangle = 0$. The $\bop_i$ are the normal master modes of the Lindblad master equation and the $\lambda_{P,i}$ the rapidities.     

\subsection{Non-positivity of the eigenvalues of the Lindbladian}

To prove the eigenvalues of $\mathcal{L}$ are non-positive, it is sufficient to prove that all the eigenvalues of the matrix $\Pm$ are non-positive. The proof is similar as in \cite{Prosen2010}. Assuming $\Pm x = \omega x$, therefore $x^{\dagger}\Pm^{\dagger} = \omega^{\ast}x^{\dagger}$, then we have
\begin{align}
x^{\dagger}(\Pm^{\dagger} + \Pm)x = x^{\dagger}\Pm^{\dagger}x + x^{\dagger}\Pm x = 2 \real(\omega)x^{\dagger}x
\end{align} 
where $\real(\omega)$ means the real part of $\omega$. 
Moreover $\Pm^{\dagger} + \Pm=\Lmp-{\Lmm}^t$ hence all the eigenvalues of the matrix on the right-hand side have to be non-positive for the master equation to have a steady state. Hence we can conclude that 
\begin{align}
\real(\omega) \leq 0,   
\end{align}
i.e. the real part of eigenvalues of the Lindbladian is non-positive.

\subsection{Computing the expectation value $\langle \aadop_i\aaop_j \rangle$}
We denote $\vert \textbf{1} \rangle = \sum_{i_1, i_2, \dots, i_L} \vert i_1, i_2, \dots, i_L, i_1, i_2, \dots, i_L \rangle$ to be the state vector resulting from the mapping of an identity operator, and $\langle \textbf{1} \vert$ to be its transpose. Computing the expectation value of observable $\hat{O}$ on the steady state $\rhoop_{ss}$, which is $\tr(\hat{O}\rhoop_{ss})$, is equivalent to the expression $\langle \textbf{1}\vert \hat{O} \vert \rho_{ss} \rangle$ where we have simply re-written the trace of an operator times the density operator in the enlarged space. Of course the operator $\hat{O}$ has also been mapped to the new space. In order to compute quadratic expectation values such as $\tr(\aadop_i\aaop_j\rhoop_{ss})=\langle \textbf{1} |\adop_i \aop_j|\rho_{ss}\rangle$ it is convenient to rewrite the eigenequation Eq.(\ref{eq:eigeneq}) in a different form. 

\subsubsection{Equation for quadratic operators}
To do so we start from $\Wm_1^{-1}\;\Wm_1 = \id_{2L}$ and using Eq.(\ref{eq:W1}), we have
\begin{align}
&\Dm = \Cm - {\Wm_P^{\dagger}}^{-1}; \\
&\Cm = \Wm_P \Qm
\end{align}
where $\Qm$ is a $L \times L$ Hermitian matrix. Therefore $\Wm_1$ can also be written as
\begin{align}
\Wm_1 = \left(
             \begin{array}{cccc}
              \Wm_P & \Wm_P \Qm  \\
              \Wm_P & \Wm_P \Qm- {\Wm_P^{\dagger}}^{-1} \\
              \end{array}
         \right) \label{eq:W1sol}
\end{align}
From Eq.(\ref{eq:eigeneq}), which can be written now as 
\begin{align}
\Zm_L \Mm \Wm_1 = \Wm_1 \left(
             \begin{array}{cccc}
              \lmp & 0  \\
              0 & -\lmpc \\
              \end{array}
         \right)
\end{align}
and using Eq.(\ref{eq:W1sol}), together with $\Omegam = \Wm_P \Qm \Wm_P^{\dagger}$ we have 
\begin{align}
\Pm \Omegam + \Omegam \Pm^{\dagger} = \Lmp \label{eq:pxxp}
\end{align}
Solving this equation for $\Omegam$ will prove very useful in the following.

\subsubsection{Relation between the elements of $\Omegam$ and $\tr(\rhoop \aadop_i \aaop_j)$}
Using Eqs.(\ref{eq:atob}) we get 
\begin{align}
\adop_i &= -\sum_{k=1}^L{\Dm}_{i,k}^{\ast}\bdop_k - \sum_{k=1}^L {\Wm_P}_{i,k}^{\ast}\bop_{L+k} \\
\aop_j &= \sum_{k=1}^L{\Wm_P}_{j,k}\bop_k + \sum_{k=1}^L \Cm_{j,k}\bdop_{L+k}
\end{align}
for $1\leq i, j \leq L$. Using this we can write 
\begin{align}
\adop_i \aop_j &= -\sum_{k,m=1}^L \Dm_{i,k}^{\ast}{\Wm_P}_{j,m} \bdop_k \bop_m - \sum_{k,m=1}^L \Dm_{i,k}^{\ast}\Cm_{j,m} \bdop_k \bdop_{L+m} \nonumber \\
 &- \sum_{k,m=1}^L {\Wm_P}_{i,k}^{\ast}{\Wm_P}_{j,m} \bop_{L+k}\bop_m \nonumber \\
 & - \sum_{k,m=1}^L {\Wm_P}_{i,k}^{\ast}\Cm_{j,m} \bop_{L+k}\bdop_{L+m} \label{eq:aiaj}
\end{align}
We then show that $\langle \textbf{1} \vert$ is annihilated by all the operators $\textbf{b}_{1 \rightarrow 2L}^{\prime}$. It is actually sufficient to prove it for all the $\textbf{b}_{1 \rightarrow L}^{\prime}$ because the $\textbf{b}_{L+1 \rightarrow 2L}^{\prime}$ have the same structure. Taking $1 \leq i \leq L$, and using Eq.(\ref{eq:btoa}), we have
\begin{align}
\langle \textbf{1} \vert b_i^{\prime} &= \sum_{i_1, i_2, \dots, i_L} \langle \textbf{1} \vert \left(\sum_{k=1}^L {\Wm_P^t}_{i,k}\adop_k - \sum_{k=1}^L {\Wm_P^t}_{i,k}\aop_{L+k}\right) \nonumber \\ 
 &= \sum_{k=1}^L {\Wm_P^t}_{i,k} \sum_{i_1, i_2, \dots, i_L} \langle \textbf{1} \vert \left(\adop_k - \aop_{L+k}\right) \nonumber \\
 &= \sum_{k=1}^L {\Wm_P^t}_{i,k} \sqrt{i_k}\sum_{i_1, i_2, \dots, i_L} \langle \dots, i_k-1, \dots, i_k, \dots \vert \nonumber \\
  &- \sum_{k=1}^L {\Wm_P^t}_{i,k} \sqrt{i_k+1}\sum_{i_1, i_2, \dots, i_L} \langle \dots, i_k, \dots, i_k+1,\dots \vert \nonumber \\
 &= \sum_{k=1}^L {\Wm_P^t}_{i,k} \sqrt{i_k+1}\sum_{i_1, i_2, \dots, i_L} \langle \dots, i_k, \dots, i_k+1, \dots \vert \nonumber \\
  &- \sum_{k=1}^L {\Wm_P^t}_{i,k} \sqrt{i_k+1}\sum_{i_1, i_2, \dots, i_L} \langle \dots, i_k, \dots, i_k+1,\dots \vert \nonumber \\ &= 0
\end{align}
Hence, computing the trace of Eq.(\ref{eq:aiaj}) we find that only the last term does not vanish and gives 
\begin{align}
\langle \textbf{1}\vert \adop_i \aop_j \vert \rho_{ss} \rangle &= -\langle \textbf{1}\vert \sum_{k,m=1}^L {\Wm_P}_{i,k}^{\ast}\Cm_{j,m} \bop_{L+k}\bdop_{L+m} \vert \rho_{ss} \rangle \nonumber \\
 &= - \sum_{k,m=1}^L {\Wm_P}_{i,k}^{\ast}\Cm_{j,m} \deltam_{k,l} 
  = -(\Cm {\Wm_P}^{\dagger})_{j,i} \nonumber \\
  &= -(\Wm_P \Qm \Wm_P^{\dagger})_{j,i} = -\Omegam_{j,i}
\end{align}
The observable matrix $\Om_{i,j} = \tr(\rhoop \aadop_i \aaop_j) = \langle \textbf{1} \vert \adop_i \aop_j \vert \rho_{ss} \rangle$ is then given by 
\begin{align}
  \Om  =   -\Omegam^t. \label{eq:OmOmega} 
\end{align}  

\section{Exact solution of a boundary driven bosonic chain}\label{sec:exact}

Here we apply our method to directly obtain the spectrum of Eq.(\ref{eq:master}) for a class of linear chains (LC) which can then be solved analytically in the limit of a long chain. 
We consider a linear lattice of $L$ sites in which each site can have identical bosons and which is driven at boundaries (similar, for example, to \cite{BermudezPlenio2012}). The Lindbladian $\mathcal{L}_{\rm LC}$ then becomes 
\begin{align}
\mathcal{L}_{\rm LC}(\rhoop)=-\frac{\im}{\hbar} [ \Hop_{\rm LC}, \rhoop ] + \mathcal{D}_{\rm LC}(\rhoop) \label{eq:masterLC} 
\end{align}
with 
\begin{align}
\Hop_{\rm LC} =-J\sum_{l=1}^{L-1}\left(\aadop_l \aaop_{l+1}+\aadop_{l+1} \aaop_{l} \right)   \label{eq:HamLC} 
\end{align}
and 
\begin{align}
\mathcal{D}_{\rm LC}(\rhoop) = \sum_{l=1,L}&\left[\Lambda^+_{l}(\aadop_{l}\rhoop \aaop_{l}-\aaop_{l}\aadop_{l}\rhoop) \right. \nonumber \\ &+ \left.  \Lambda^-_{l}(\aaop_{l}\rhoop \aadop_{l}-\aadop_{l}\aaop_{l}\rhoop)+ \hc \right], \label{eq:DissLC}    
\end{align}
where $\Lambda^+_l$ and $\Lambda^-_l$ are respectively the raising and lowering rates at site $l$, while $J$ is the tunnelling amplitude.  

In this case, the only non-zero elements of the matrix $\hm$ are
\begin{eqnarray}
\hm_{j, j+1} = \hm_{j+1, j} = -J.  
\end{eqnarray}
For the dissipation we rewrite the four coefficients $\Lambda^a_l$ with four new parameters   
\begin{eqnarray}
&&\Gamma_1 =  \Lambda_1^- - \Lambda^+_1, \;\; \bar{n}_1 = \frac{\Lambda_1^+}{\Gamma_1} \\
&&\Gamma_L =  \Lambda_L^- - \Lambda^+_L, \;\; \bar{n}_L = \frac{\Lambda_L^+}{\Gamma_L}
\end{eqnarray}
Therefore, all the non-zero elements of matrix $\Pm$ are
\begin{align}
&\Pm_{1,1} = -\frac{\Gamma_1}{2}, \;\; \Pm_{L,L} = - \frac{\Gamma_L}{2} \\
&\Pm_{m, m+1} = \Pm_{m+1,m} =  \frac{\im J}{2 \hbar}   
\end{align}
for $ 1 \leq m < L$. Note that $\bar{n}_1$ and $\bar{n}_L$ do not appear in the matrix $\Pm$, and hence will not affect the rapidities.  

\begin{figure}
\includegraphics[width=\columnwidth]{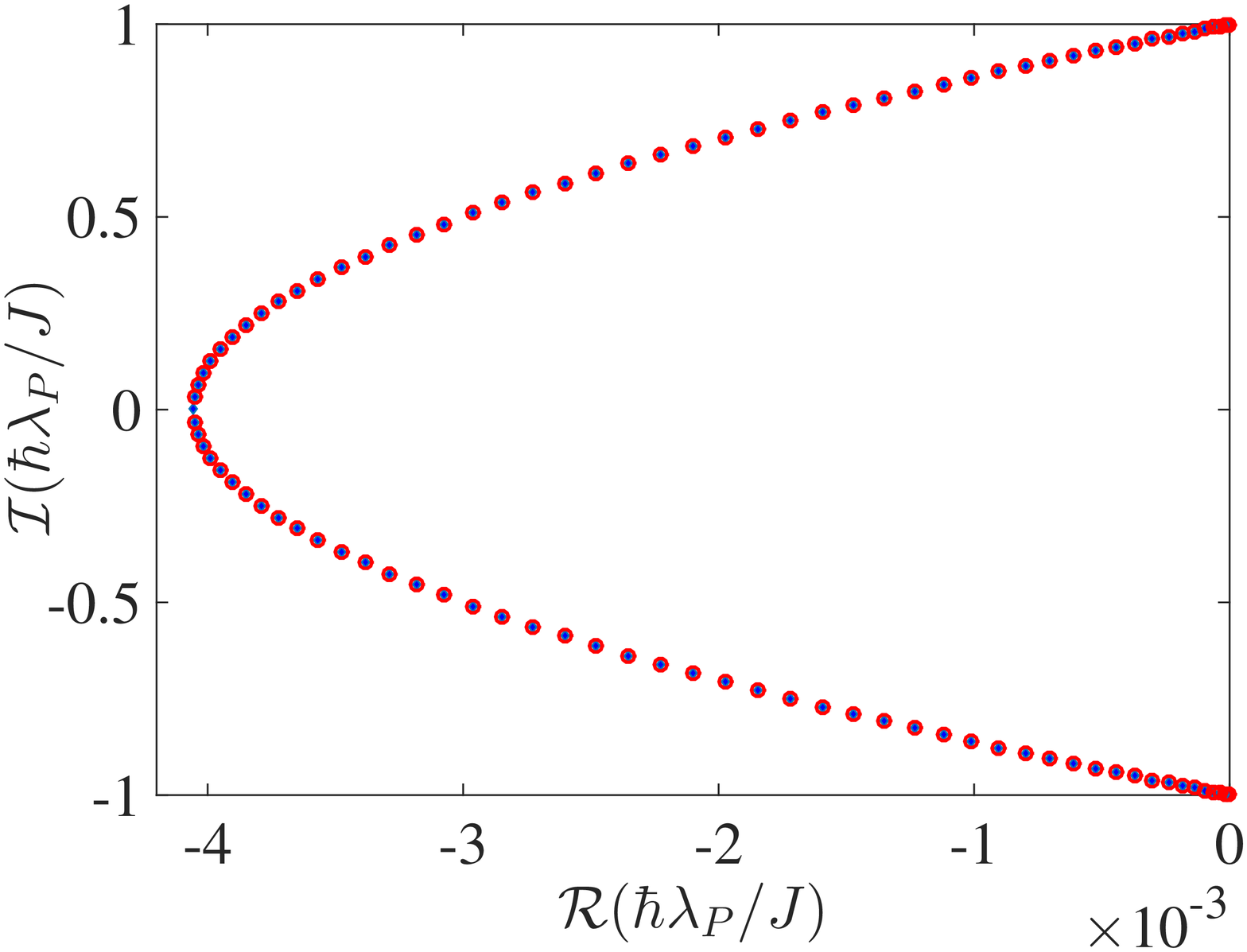}
\caption{(color online) Eigenvalues $\lmp$ of $\Pm$ which correspond to the spectrum of the Lindbladian $\mathcal{L}_{\rm LC}$ of Eqs.(\ref{eq:masterLC}-\ref{eq:DissLC}), for a boundary driven bosonic chain of length $L=100$, $\hbar\Gamma_1/J = 5$, $\hbar\Gamma_L/J = 1/5$. The blue diamonds are given by the numerical solution of Eq.(\ref{eq:eigenP}), while the red circles are given by the analytical solution Eq.(\ref{eq:lambda}) using Eqs.(\ref{eq:alpha},\ref{eq:beta}). $\real(\omega)$ and $\imag(\omega)$ mean respectively the real and the imaginary part of the complex number $\omega$. \label{fig:spectrum} }
\end{figure}

This results in the important fact that $\Pm$ is a tridiagonal matrix whose elements are constant along the diagonals (i.e. Toeplitz) except for top-left and bottom-right corners (i.e. bordered). The eigenvalues and eigenvectors of this matrix can be analytically computed \cite{Yueh2005} (for the eigendecomposition of more general tridiagonal matrices see for example \cite{Kouachi2006, Willms2008}). Assuming that $\lambda$ is an eigenvalue of $\Pm$, and $u$ is the corresponding right eigenvector so that $\Pm u = \lambda u$, then we find that $\lambda$ and $u$ are given by 
\begin{align}  
\lambda = \im \frac{J}{\hbar} \cos \pare{\theta} \label{eq:lambdaexact}
\end{align}    
and the $L$ elements of $u$ are   
\begin{align}
u_j = \frac{u_1}{\sin \pare{\theta}} \left\{\sin \pare{j\theta} - \im \frac{\hbar\Gamma_1}{J}\sin[(j-1)\theta] \right\}.  \label{eq:u}   
\end{align}
In Eqs.(\ref{eq:lambdaexact},\ref{eq:u}) $\theta$ is a complex number which satisfies the equality     
\begin{align}
& -\frac{J^2}{\hbar^2} \sin[(L+1)\theta] + \im \frac{J}{\hbar}(\Gamma_1 + \Gamma_L)\sin \pare{L\theta} \nonumber \\
& + \Gamma_1 \Gamma_L\sin\spare{(L-1)\theta} = 0  
\end{align}
except the trivial solutions $\theta \neq m\pi$ with $ m \in Z$. 

Denoting $\theta = \alpha + \im\beta $, we transform the above equation into two equations of real numbers 
\begin{align}
& \frac{J^2}{\hbar^2} \sin[(L+1)\alpha]\cosh[(L+1)\beta]  \nonumber \\ 
&+ \frac{J}{\hbar} (\Gamma_1 + \Gamma_L)\cos \pare{L\alpha} \sinh (L\beta) \nonumber \\
&-\Gamma_1 \Gamma_L \sin[(L-1)\alpha] \cosh[(L-1)\beta] = 0 \label{eq:real1} \\
& \frac{J^2}{\hbar^2}\cos[(L+1)\alpha] \sinh[(L+1)\beta] \nonumber \\ &- \frac{J}{\hbar} (\Gamma_1 + \Gamma_L)\sin \pare{L\alpha} \cosh (L\beta) \nonumber \\ 
& - \Gamma_1 \Gamma_L\cos[(L-1)\alpha] \sinh[(L-1)\beta] = 0 \label{eq:real2} 
\end{align}
In the following we solve the above equation approximately, in the limit $L \rightarrow \infty$. All the solutions are found for $\alpha \in [0, \pi]$ \cite{Willms2008}. We then make the approximation that
\begin{align}
& \sinh[(L+1)\beta] \simeq \sinh[(L-1)\beta] \simeq \sinh [L\beta] \\
& \cosh[(L+1)\beta] \simeq \cosh[(L-1)\beta] \simeq \cosh [L\beta] 
\end{align}
with which Eqs.(\ref{eq:real1},\ref{eq:real2}) become 
\begin{align}
&\frac{J^2}{\hbar^2}\sin\spare{(L+1)\alpha} + \frac{J}{\hbar}(\Gamma_1 + \Gamma_L)\cos \pare{L\alpha} \tanh \pare{L\beta} \nonumber \\
&- \Gamma_1 \Gamma_L \sin\spare{(L-1)\alpha} = 0 \label{eq:ab1}\\
&\frac{J^2}{\hbar^2} \cos\spare{(L+1)\alpha} \tanh \pare{L\beta} - \frac{J}{\hbar}(\Gamma_1 + \Gamma_L)\sin \pare{L\alpha} \nonumber \\
&- \Gamma_1 \Gamma_L\cos\spare{(L-1)\alpha} \tanh \pare{L\beta} = 0   \label{eq:ab2}       
\end{align}
Combining Eqs.(\ref{eq:ab1},\ref{eq:ab2}) we get
\begin{align}
& \frac{ J^2\sin\spare{(L+1)\alpha} - \hbar^2 \Gamma_1 \Gamma_L \sin[(L-1)\alpha]}{\hbar J(\Gamma_1 + \Gamma_L)\cos \pare{L\alpha}} \nonumber \\
= & \frac{\hbar J(\Gamma_1 + \Gamma_L)\sin \pare{L\alpha}}{ \hbar^2\Gamma_1 \Gamma_L \cos\spare{(L-1)\alpha} - J^2\cos\spare{(L+1)\alpha} }
\end{align}
which can be rewritten as  
\begin{align}
& (\kappa_1 + \kappa_L)\sin \pare{2L\alpha} + \sin \spare{2(L-1)\alpha} \nonumber \\
& + \kappa_1 \kappa_L \sin \spare{2(L+1)\alpha} = 0 \label{eq:ak}
\end{align}
where $\kappa_1 = J^2/(\hbar^2\Gamma_1^2)$ and $\kappa_L = J^2/(\hbar^2\Gamma_L^2)$. Eq.(\ref{eq:ak}) can be solved analytically when $J^2 =\hbar^2 \Gamma_1 \Gamma_L$. In fact $\kappa_1 = \frac{1}{\kappa_L} = \kappa$ and Eq.(\ref{eq:ak}) reduces to
\begin{align}
\left(\kappa + \frac{1}{\kappa} \right)\sin \pare{2L\alpha} + \sin \spare{2(L-1)\alpha} + \sin \spare{2(L+1)\alpha} = 0 \label{eq:simpsol}
\end{align}
or equivalently $\spare{\kappa + \frac{1}{\kappa} + 2\cos \pare{\alpha}}\sin \pare{2L\alpha} = 0$. Since $\kappa + 1/\kappa    \ge 2$, 
the real solutions are 
\begin{align}
\alpha = \frac{k\pi}{L} \label{eq:alpha}
\end{align}     
with $1 \leq k < L$. Note that the solutions of Eq.(\ref{eq:simpsol}) $\alpha = k\pi/2L$ with $k$ odd have been discarded because inconsistent with Eqs.(\ref{eq:ab1},\ref{eq:ab2}). 

We then get for $\beta$ 
\begin{align}
\tanh \pare{L\beta} = -\frac{2 \sqrt{\kappa}}{\kappa + 1} \sin \frac{k\pi}{L},
\end{align}
which results in      
\begin{align}
\beta = \frac{1}{2L}\ln\left(\frac{1-  \frac{2 \sqrt{\kappa}}{\kappa + 1}\sin \frac{k \pi}{L}}{1 +  \frac{2 \sqrt{\kappa}}{\kappa + 1}\sin \frac{k\pi}{L}} \right) \label{eq:beta}
\end{align}
Using Eqs.(\ref{eq:lambdaexact}) and that $\theta=\alpha+\im\beta$ we have
\begin{align}
\lambda = \frac{J}{\hbar}\sin \pare{\alpha} \sinh \pare{\beta} + \im \frac{J}{\hbar}\cos \pare{\alpha} \cosh \pare{\beta}. \label{eq:lambda}
\end{align} 
Inserting the various $\alpha$ and $\beta$ from Eqs.(\ref{eq:alpha},\ref{eq:beta}) we have an analytical solution for the eigenvalues. 
In Fig.\ref{fig:spectrum} we compare the analytical solution Eq.(\ref{eq:lambda}), blue diamonds, with numerical evaluation of the spectrum for Eq.(\ref{eq:masterLC}), red circles, for a system of length $L=100$. The figure shows a remarkable match of the analytical and numerical spectra.    

A natural consequence of knowing the spectrum, is that it is possible to compute the relaxation gap $\Delta$, which is given by the eigenvalues of the Linbladian with the real part closest to zero. We thus get     
\begin{align}
\Delta &= \frac{J}{\hbar}\sin \pare{\frac{\pi}{L}} \sinh \left[ \frac{1}{2L}\ln \left(\frac{1 +  \frac{2 \sqrt{\kappa}}{\kappa + 1}\sin \pare{\frac{\pi}{L}}}{1 -  \frac{2 \sqrt{\kappa}}{\kappa + 1}\sin \pare{\frac{\pi}{L}}} \right) \right] \\ \nonumber
&\simeq  \frac{2 \sqrt{\kappa}J}{\hbar(\kappa + 1)L} \sin^2 \pare{\frac{\pi}{L}} \\ \nonumber &\simeq  \frac{2\pi^2 \sqrt{\kappa}}{\hbar(\kappa + 1)} \frac{J}{L^3},
\end{align}
which scales as $1/L^3$, in agreement with the predictions in \cite{Prosen2008, Znidaric2015}.

\section{Efficient algorithm to compute quadratic observables}\label{sec:algorithm}
Eq.(\ref{eq:pxxp}) is a Lyapunov equation, which is a special case of Sylvester equation \cite{BartelsStewart1972, GolubLoan1979}. There exists numerical methods to efficiently solve this type of equations of order $O(L^3)$. In this section we also propose an $O(L^3)$ algorithm to solve Eq.(\ref{eq:pxxp}) based on findings in the previous sections. We start by solving the eigenvalue decomposition problem of matrix $\Pm$, to get the left eigenvector space $\Wm_P^l$, the right eigenvector space $\Wm_P$ and the diagonal matrix of eigenvalues $\lmp=\diag(\lambda_{P,1},\dots \lambda_{P,L})$. At this point it is already possible to figure out whether the system has any dark mode. In fact this would be manifested by the existence of eigenvalues with zero real part. If the real parts of all eigenvalues of $\Pm$ are strictly smaller than $0$, then the system has no dark modes, the steady state is unique and the following algorithm can be used. We can thus write $\Omegam=\Wm_P \Qm \Wm_P^{\dagger}$ which gives 
\begin{align}\label{eq:forQ}  
& \Pm \Omegam + \Omegam \Pm^{\dagger} \\ \nonumber = & \Pm \Wm_P \Qm \Wm_P^{\dagger} + \Wm_P \Qm \Wm_P^{\dagger} \Pm^{\dagger} \\ \nonumber = & \Wm_P \lmp \Qm \Wm_P^{\dagger} + \Wm_P \Qm \lmp^{\ast} \Wm_P^{\dagger} \\ \nonumber = &\Lmp    
\end{align}
We can renormalize $\Wm_P$ and $\Wm_P^l$ so that $\Wm_P^l = \Wm_P^{-1}$. From Eq.(\ref{eq:forQ}) we get the elements of the Hermitian matrix $\Qm$ by
\begin{eqnarray} \label{eqofQ}
&& \lmp \Qm + \Qm \lmpc = \Wm_P^{-1}\Lmp (\Wm_P^{-1})^{\dagger} = \Wm_P^l\Lmp {\Wm_P^l}^{\dagger} \nonumber \\
&& \Leftrightarrow \Qm_{m,n} = \frac{( \Wm_P^l\Lmp {\Wm_P^l}^{\dagger} )_{m,n}}{\lambda_{P,m} + \lambda_{P,n}^{\ast}} \label{eq:solveQ} 
\end{eqnarray} 

The elements of the matrix corresponding to $\langle\aadop_i \aaop_j\rangle$, i.e. $\Om_{i,j} = \tr(\rhoop \aadop_i \aaop_j)$ are then given by, using Eq.(\ref{eq:OmOmega})       
\begin{align}
\Om_{i,j} = - \Omegam_{j,i} = -\sum_{m,n} \Wm_{P j,m} \Qm_{m,n} \Wm^*_{P i,n} \label{eq:Oij}   
\end{align}

Note that with this approach we only need to solve an eigenvalue decomposition problem, plus a few matrix multiplications. All the matrices involved in these procedures are of size $L \times L$. The complexity of this algorithm is thus $O(L^3)$ which is the complexity of solving a $L \times L$ non-Hermitian eigendecomposition problem. 

It should also be noted that because of the structure of Eq.(\ref{eq:solveQ}), this approach can become unstable if $\Pm$ has eigenvalues whose real part is very close to $0$.

\subsection{Scaling of the relaxation time across the phase transition}

\begin{figure}
\includegraphics[width=0.9\columnwidth]{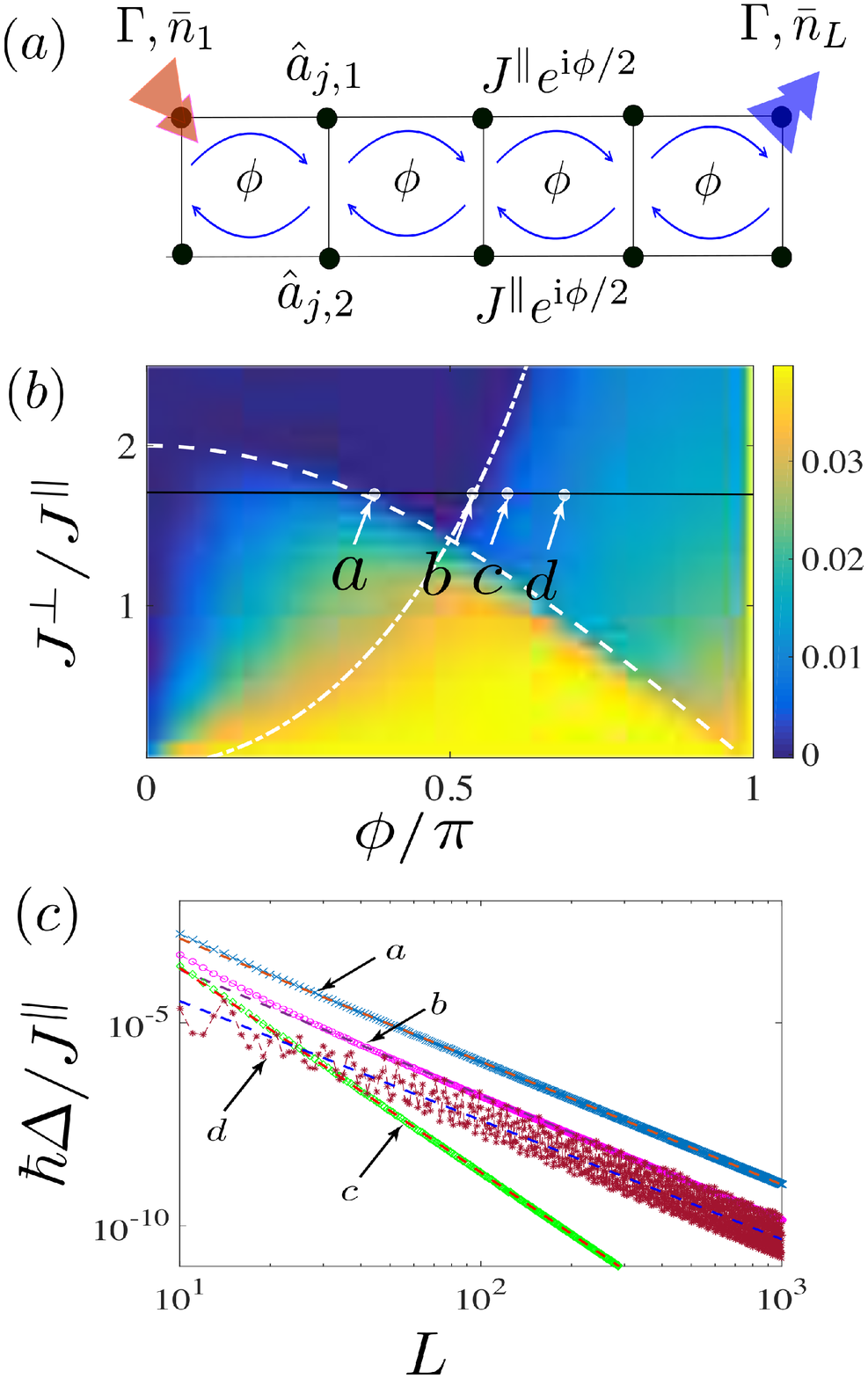}
\caption{(color online) (a) Ladder made of two coupled linear chains, with local bosonic excitations described by the annihilation operators at site $j$, $\aaop_{j,p}$, where $p = 1,2$ for the upper and the lower leg respectively. $\Jp$ is the tunnelling between the legs, while $\Jpa$ is the tunnelling between sites in the legs. A gauge field imposes a phase $\phi$. (b) Chiral current $\mathcal{J}_c$ as a function of $\Jp$ and $\phi$ for $L = 500$. The white dashed and the white dot-dashed lines correspond respectively to the two phase transitions in Eqs.(\ref{eq:totcurtransition}, \ref{eq:qptransition}) respectively. The black horizonal line corresponds to the line $\Jp = 1.7$, and the four white circles a, b, c and d on this line correspond to $\phi = \phi_{c1}, 0.5, \phi_{c1}, 0.6$ respectively. Panels (a) and (b) are similar to \cite{GuoPoletti2016}. (c) The relaxation time $\hbar\Delta/\Jpa$ versus the length of the ladder $L$. Both $\hbar\Delta/\Jpa$ and $L$ are shown in log scale so that an algebraic decay is clearly represented by a straight line. The four lines correspond to the point a, b, c and d in panel (b). The line marked with green diamonds (line c) shows the scaling at $\phi = 0.5398$ which corresponds to the non-equilibrium phase transition described by Eq.(\ref{eq:qptransition}). The red dashed line is a linear fitting of line c, which has the exponent $-5$. The other three straight dashed lines are linear fittings of a, b and d, all with the same exponent $-3$. \label{fig:gap} }
\end{figure}

With the algorithm introduced by Eqs.(\ref{eq:solveQ},\ref{eq:Oij}), we can more effectively explore larger systems, allowing to easily study the scaling of the relaxation gap in various non-equilibrium phases. In the following we apply our method to study the boundary driven bosonic ladder when also a magnetic field is imposed on it. This system is depicted in Fig.\ref{fig:gap}(a). The Lindblad master equation for this bosonic ladder (BL) is 
\begin{align}
\mathcal{L}_{\rm BL}(\rhoop) = -\frac{\im}{\hbar} \left[ \Hop_{\rm BL}, \rhoop \right] + \mathcal{D}_{\rm BL}(\rhoop), \label{eq:masterBL}   
\end{align}
with Hamiltonian $\Hop_{\rm BL}$ given by 
\begin{align}
\Hop_{\rm BL}=-&\left(J^{\parallel} \sum_{p,j}\right.\allowbreak e^{\im(-1)^{p+1}\phi/2}\aadop_{j,p}\aaop_{j+1,p} \nonumber\\
&+ \allowbreak\left.\Jp \sum_{j} \aadop_{j,1}\aaop_{j,2} \right) + {\rm H.c.} 
\end{align}
Here $J^{\parallel}$ is the tunnelling constant in the legs, and $\Jp$ for the rungs. $\aaop_{j,p}$ $(\aadop_{j,p})$ annihilates (creates) a boson in the upper (for $p=1$) or lower (for $p=2$) chain at the $j$-th rung of the ladder. A particle tunneling around a plaquette would acquire a net phase of $\phi$. We consider a dissipative coupling $\Dop_{\rm BL}$ on the two edges modelled by
\begin{align} 
\Dop_{\rm BL}(\Op)  = \sum_{j=1,L} \Gamma \allowbreak\left[ \allowbreak\bar{n}_{j,1} \left(\aaop_{j,1}\Op\aadop_{j,1} - \aaop_{j,1}\aadop_{j,1}\Op \right) \right. \nonumber \\
\allowbreak+  \left. (\bar{n}_{j,1} +1)\right. \allowbreak  \left.\left(\aadop_{j,1}\Op\aaop_{j,1} - \aadop_{j,1}\aaop_{j,1}\Op \right) \allowbreak + {\rm H.c.} \right]  
\end{align}
where $\Gamma$ is the coupling constant of the bosons at sites $j=1,L$, while $\bar{n}_{j,1}$ is the local particle density that the dissipator would impose to the bosonic site if the site was isolated. The dissipator is only coupled to the sites at the extremeties of the upper leg [see Fig.\ref{fig:gap}(a)]. 

The unitary counterpart of this system is known to exhibit a quantum phase transition from the Meissner to the vortex phase. The transition between the two phases is characterized by the chiral current $\mathcal{J}_c$, defined as the difference of the currents between the upper and the lower leg
\begin{align}
\mathcal{J}_c = \sum_j (\mathcal{J}_{j,1} - \mathcal{J}_{j,2})/L,
\end{align}
where $\mathcal{J}_{j,p} = \langle \im \Jpa e^{\im (-1)^{p+1} \phi/2}\aadop_{j,p}\aaop_{j+1,p} + {\rm H.c} \rangle$ is the particle current out of site $j$ on the $p$-th leg. In the Meissner phase, $\mathcal{J}_c$ is non-zero, while in the vortex phase, $\mathcal{J}_c$ is greatly suppressed \cite{Kardar1986, Granato1990, DennistonTang1995}. For an experimental realization with ultracold gases see \cite{AtalaBloch2014}. 

For the open case this system was studied in detail in \cite{GuoPoletti2016} where it was shown that two non-equilibrium phase transitions can emerge between phases with or without chiral current. Moreover, one of the two transitions would also be signalled by a sudden suppression of the current. The coupling to the baths studied here corresponds to the $R$ configuration of \cite{GuoPoletti2016} for which these two transitions can occur. 

Translating this model to the elements of $\Pm$ we get 
%
%
\begin{align}
\Pm_{(1,1),(1,1)} &= \Pm_{(L,1), (L,1)} = - \Gamma \\
\Pm_{(j, 1),(j,2)} &= \Pm_{(j,2),(j,1)} = \im \frac{\Jp}{2\hbar} \\
\Pm_{(j,p),(j+1,p)} &= \Pm_{(j+1,p),(j,p)}^{\ast} = \im \frac{\Jpa}{2\hbar} e^{\im (-1)^{p+1}\phi/2} \label{eq:PJpa}       
\end{align}
for $ 1 \leq j \leq L$ (except in Eq.(\ref{eq:PJpa}) for which $j < L$). All the other elements of $\Pm$ are zero. We note that in this case $\Pm$ is a block bordered Toeplitz matrix for which, to the best of our knowledge, the analytical eigendecomposition is not known \cite{Prosen2008}. The two phase transitions occur, respectively, for $\phi=\tilde{\phi},\;\bar{\phi}$ and $\Jp=\tilde{J}^{\perp},\;\bar{J}^{\perp}$ given by    
\begin{align}
\bar{J}_{\perp}&=2\Jpa\cos\left(\bar{\phi}/2\right)  \label{eq:totcurtransition} \\
\tilde{J}_{\perp}&=2\Jpa\tan\left(\tilde{\phi}/2\right)\sin\left(\tilde{\phi}/2\right) \label{eq:qptransition}   
\end{align}
as shown in Fig.\ref{fig:gap}(b). The transition line Eq.(\ref{eq:totcurtransition}) is depicted by a white dashed line, while the other transition line, Eq.(\ref{eq:qptransition}), is represented by a white dot-dashed line.   

Here we focus on the scaling of the relaxation gap of the Linbladian (\ref{eq:masterBL}) across the two open quantum phase transitions. In Fig.(\ref{fig:gap}), we have chosen $J^{\perp}/\Jpa = 1.7$, where the system exhibits two phase transitions at $\phi=\phi_{c1} \approx 0.3532$ and $\phi=\phi_{c2} \approx 0.5398$, calculated from Eqs.(\ref{eq:totcurtransition}) and (\ref{eq:qptransition}). 
In Fig.\ref{fig:gap}(c) we show the scaling of the relaxation gap as the size of the system increases (we consider $L=10\rightarrow 1000$). The gap is analyzed in $4$ distinct points $a$, $b$, $c$ and $d$ for $\phi=\phi_{c1},\;0.5,\;\phi_{c2}$, and $0.6 $ as shown by the white dots in Fig.\ref{fig:gap}(b). For the parameters corresponding to points $a$, $b$ and $d$ the scaling of the gap is proportional to $L^{-3}$ as shown by, respectively, the blue crosses, the pink circles and the red stars in Fig.\ref{fig:gap}(c). All the fits are represented by dashed lines.   
For the transition point $c$, green diamonds, the scaling is instead $L^{-5}$ as predicted in \cite{Prosen2008, Znidaric2015}. It is here important to discuss the difference in scaling of the relaxation gap in the two transition lines. For the line given by Eq.(\ref{eq:qptransition}), and hence also point $c$, the Hamiltonian of the bulk system presents a quantum phase transition, and the energy spectrum goes from one to two minima. At the transition point the spectrum is not quadratic but quartic thus affecting the scaling of the relaxation gap. Instead, for the line given by Eq.(\ref{eq:totcurtransition}) the low energy spectrum is not qualitatively changed, instead a gap opens (see \cite{GuoPoletti2016}). While the opening of the gap affects the total and chiral currents, it does not change the scaling of the relaxation gap. At the intersection point between the two transition lines the scaling is indeed $L^{-5}$.

\section{Computing the steady state}\label{sec:steady_state}

From Eq.(\ref{eq:Llambdap}) we understand that the steady state of the system is the vacuum of the operators $\bop_j$, that is $|\rho_{ss}\rangle=|\zero \rangle_b$. This is related to the vacuum of the $\aop_j$, $|\zero \rangle_a$, by a linear transformation. We can then write 
\begin{align} \label{eq:steadystate}
\vert \rho_{ss} \rangle = \Sop^{-1} \vert \zero \rangle_a.  
\end{align}       

In the following we show how to compute $\Sop$ from $\Wm_1$. First we write $\Sop = e^{\Top}$, where $\Top$ is
 \begin{align}
\Top &= \frac{1}{2} \left(
   \begin{array}{cc}
  \textbf{a}^{\dagger}_{1\rightarrow L} \\
   \textbf{a}_{L+1\rightarrow 2L} \\
   \end{array}
  \right)^t \left(
             \begin{array}{cccc}
              \Um & \Vm  \\
              \IM & \Jm \\
              \end{array}
         \right)
  \left(
      \begin{array}{cc}
   \textbf{a}_{1\rightarrow L} \\
  \textbf{a}^{\dagger}_{L+1\rightarrow 2L} \\
   \end{array}
    \right)
 \nonumber \\ 
 &+ \frac{1}{2} \left(
      \begin{array}{cc}
   \textbf{a}_{1\rightarrow L} \\
  \textbf{a}^{\dagger}_{L+1\rightarrow 2L} \\
   \end{array}
    \right)^t \left(
             \begin{array}{cccc}
              \Um^t & \IM^t  \\
              \Vm^t & \Jm^t \\
              \end{array}
         \right)
  \left(
   \begin{array}{cc}
  \textbf{a}^{\dagger}_{1\rightarrow L} \\
   \textbf{a}_{L+1\rightarrow 2L} \\
   \end{array}
  \right) \label{eq:Top}
\end{align}
where $\Um, \Vm, \IM, \Jm$ are $L\times L$ matrix. Hereafter we will write
\begin{align}
\Wm = \left(
             \begin{array}{cccc}
              \Um & \Vm  \\
              \IM & \Jm \\
              \end{array}
         \right)
\end{align}
To calculate $e^{\Top}\adop_j e^{-\Top}$ and $e^{\Top}\aop_{L+j} e^{-\Top}$, we use the relations 
\begin{align}
&\hat{E}  := e^{\hat{T}}\adop_j e^{-\hat{T}} = \sum_{m=1}^{\infty}\frac{1}{m!}\left[\Top,\adop_j\right]_m \nonumber \\
&\hat{F} :=  e^{\hat{T}}\aop_{L+j} e^{-\hat{T}} = \sum_{m=1}^{\infty}\frac{1}{m!}\left[\Top,\aop_{L+j}\right]_m  \nonumber
\end{align} 
where the nested commutator is defined recursively as $[\hat{A},\hat{B}]_{m+1} \equiv [\hat{A},[\hat{A},\hat{B}]_{m}] $ with $[\hat{A},\hat{B}]_0 \equiv \hat{B}$. 
After a little algebra it is possible to show that
\begin{eqnarray}
&& \Sop \left(
   \begin{array}{cc}
  \textbf{a}^{\dagger}_{1\rightarrow L} \\
   \textbf{a}_{L+1\rightarrow 2L} \\
   \end{array}
  \right)^t \Sop^{-1} \\ \nonumber 
         = &&\left(
   \begin{array}{cc}
  \textbf{a}^{\dagger}_{1\rightarrow L} \\
   \textbf{a}_{L+1\rightarrow 2L} \\
   \end{array}
  \right)^t e^{\Wm \Zm_L}
\end{eqnarray}
Similarly, we have
\begin{eqnarray}
\Sop \left(
      \begin{array}{cc}
   \textbf{a}_{1\rightarrow L} \\
  \textbf{a}^{\dagger}_{L+1\rightarrow 2L} \\
   \end{array}
    \right) \Sop^{-1} = e^{-\Zm_L \Wm} \left(
      \begin{array}{cc}
   \textbf{a}_{1\rightarrow L} \\
  \textbf{a}^{\dagger}_{L+1\rightarrow 2L} \\
   \end{array}
    \right)
\end{eqnarray}
We now denote $\tilde{\Wm}_1 = e^{-\Zm_L \Wm}, \tilde{\Wm}_2 = e^{\Wm \Zm_L}$, and we see that
\begin{eqnarray}
\log\left(\tilde{\Wm}_1\right) \Zm_L + \Zm_L \log\left(\tilde{\Wm}_2\right) = 0
\end{eqnarray}
which allows us to write       
\begin{eqnarray}
\tilde{\Wm}_2 = \Zm_L \tilde{\Wm}_1^{-1} \Zm_L
\end{eqnarray}
Therefore we have that 
\begin{align}\label{eq:SLS}
& \Sop \mathcal{L} \Sop^{-1} \nonumber \\ = & \left(
   \begin{array}{cc}
  \textbf{a}^{\dagger}_{1\rightarrow L} \\
   \textbf{a}_{L+1\rightarrow 2L} \\
   \end{array}
  \right)^t \tilde{\Wm}_2 \Mm
           \tilde{\Wm}_1  \left(
     \begin{array}{cc}
     \textbf{a}_{1\rightarrow L} \\
      \textbf{a}^{\dagger}_{L+1\rightarrow 2L} \\
      \end{array}
          \right)  \nonumber \\
+ & \left(
     \begin{array}{cc}
     \textbf{a}_{1\rightarrow L} \\
      \textbf{a}^{\dagger}_{L+1\rightarrow 2L} \\
      \end{array}
          \right)^t \tilde{\Wm}_1^t \Mm^t \tilde{\Wm}_2^t
             \left(
        \begin{array}{cc}
      \textbf{a}^{\dagger}_{1\rightarrow L} \\
     \textbf{a}_{L+1\rightarrow 2L} \\
    \end{array}
    \right)
   \nonumber \\ 
   + & \tr({\Lmm}^t-\Lmp) \nonumber \\    
    = & \left(
      \begin{array}{cc}
      \textbf{a}^{\dagger}_{1\rightarrow L} \\
    \textbf{a}_{L+1\rightarrow 2L} \\
    \end{array}
    \right)^t \Zm_L \tilde{\Wm}_1^{-1} \Zm_L \Mm
           \tilde{\Wm}_1  \left(
     \begin{array}{cc}
     \textbf{a}_{1\rightarrow L} \\
      \textbf{a}^{\dagger}_{L+1\rightarrow 2L} \\
      \end{array}
          \right)  \nonumber \\
+ & \left(
     \begin{array}{cc}
     \textbf{a}_{1\rightarrow L} \\
      \textbf{a}^{\dagger}_{L+1\rightarrow 2L} \\
      \end{array}
          \right)^t \tilde{\Wm}_1^t \Mm^t \Zm_L (\tilde{\Wm}_1^t)^{-1} \Zm_L
             \left(
      \begin{array}{cc}
      \textbf{a}^{\dagger}_{1\rightarrow L} \\
    \textbf{a}_{L+1\rightarrow 2L} \\
    \end{array}
    \right)
   \nonumber \\ 
   + & \tr({\Lmm}^t-\Lmp)
\end{align}
Thus we see that if we set
$\tilde{\Wm}_1 = \Wm_1$
which means  
\begin{eqnarray} \label{eq:logofeigenspace}
\Wm &=& -\Zm_L \log{\Wm_1},
\end{eqnarray}
then following from Eq.(\ref{eq:me_beta}) and the explicit construction of $\Wm_1$ in Sec.\ref{sec:diagonalization}, Eq.(\ref{eq:SLS}) can be simply written as
\begin{align}
 \Sop \mathcal{L} \Sop^{-1} = 2\sum_{i=1}^{L}\lambda_{P,i}\adop_{i}\aop_{i} + 2\sum_{i=1}^L \lambda_{P,i}^{\ast}\adop_{L+i}\aop_{L+i}.
\end{align}
 It follows that the vacuum $\vert \zero \rangle_a$ is the steady state of $\Sop \mathcal{L} \Sop^{-1}$ which implies that the steady state of $\mathcal{L}$ is given by Eq.(\ref{eq:steadystate}). Since $\Wm_1$ is given by Eq.(\ref{eq:logofeigenspace}), we can also reconstruct $\Top$ from Eq.(\ref{eq:Top}).

\section{Conclusions}\label{sec:conclusion} 

We have shown how to map the problem of computing the relaxation rates and the normal master modes of a Lindblad master equation for dissipatively boundary driven uniform non-interacting bosons chain, to the diagonalization of a tridiagonal bordered Toeplitz matrix. This special structure of the matrix also allows to find explicit analytical solutions and we have shown an approximate solution for a large system. With the approach presented, for a system of size $L$, the matrix to be diagonalized is only of size $L\times L$ (when considering Hamiltonians which conserve the total number of particles, i.e. there are no terms of the type $\aaop_i\aaop_j$ or $\aadop_i\aadop_j$). For more general Hamiltonians our approach can be readily extended, however the matrix to be diagonalized would be a $2L \times 2L$ block bordered Toeplitz matrix (for uniform bulk Hamiltonian with boundary dissipative driving) which cannot be diagonalized with the same analytical formulae. 
The method here presented can be useful to study both the time evolution (since it gives access to all the normal master modes and rapidities) and steady states for open bosonic systems far from equilibrium. Due to its simplicity, this method allows one to find more analytically solvable solutions.

We have also proposed a numerical algorithm which can efficiently compute observables of the type $\langle \aadop_i \aaop_j \rangle$. We have then used this method to compute the relaxation gap of a boundary driven bosonic quadratic system which presents two different non-equilibrium phase transitions. A scaling analysis of the gap shows that the gap scale as $1/L^3$ in all the parameter space except at one of the two phase transitions, for which the relaxation gap scales as $1/L^5$. This is due to the different behavior of the spectrum of the Hamiltonian of the bulk of the system at the two transition points.    

In the second part of our work, we are going to extend this approach to fermionic systems.

\begin{acknowledgments} 
We acknowledge insightful discussions with U. Bissbort. D.P. acknowledges support from Singapore Ministry
of Education, Singapore Academic Research Fund Tier-I (project SUTDT12015005).
\end{acknowledgments}

\end{document}